\documentclass{emulateapj}

\received{2010 December 14}
\slugcomment{To appear in {\it the Astrophysical Journal, Letter}}

\shortauthors{KOBAYASHI et al.}
\shorttitle{Chemical Enrichment in the Carbon-enhanced Damped Lyman $\alpha$ System}

\def\gtsim {>\kern-1.2em\lower1.1ex\hbox{$\sim$}~}   
\def\ltsim {<\kern-1.2em\lower1.1ex\hbox{$\sim$}~}   

\begin{document}

\title{Chemical Enrichment in the Carbon-enhanced Damped Lyman $\alpha$ System by the Population III Supernovae}
\author{Chiaki KOBAYASHI$^{1,3}$, Nozomu TOMINAGA$^{2,3}$,
and Ken'ichi NOMOTO$^3$}
\affil{$^1$ Research School of Astronomy \& Astrophysics, The Australian National University, Cotter Rd., Weston CT 2611, Australia; chiaki@mso.anu.edu.au}
\affil{$^2$ Department of Physics, Faculty of Science and Engineering, Konan University, 8-9-1 Okamoto, Kobe, Hyogo 658-8501, Japan}
\affil{$^3$ Institute for the Physics and Mathematics of the Universe, University of Tokyo, Kashiwa, Chiba 277-8583, Japan}

\begin{abstract}
We show that the recently observed elemental abundance pattern of the carbon-rich metal-poor Damped Lyman $\alpha$ (DLA) system is in excellent agreement with the nucleosynthesis yields of faint core-collapse supernovae of primordial stars.
The observed abundance pattern is not consistent with the nucleosynthesis yields of pair-instability supernovae.
The DLA abundance pattern is very similar to that of carbon-rich extremely metal-poor (EMP) stars, and the contributions from low-mass stars and/or binary effects should be very small in DLAs.
This suggests that chemical enrichment by the first stars in the first galaxies is driven by core-collapse supernovae from $\sim 20-50 M_\odot$ stars, and also supports the supernova scenario as the enrichment source of EMP stars in the Milky Way Galaxy.
\end{abstract}


\section{Introduction}
At the end of the dark age of the Universe, 
the cosmic dawn was heralded by the birth of the first stars and galaxies.
The nature of these first objects is still
far from being well understood.
From the theory of star formation from primordial gas, the first stars are believed to be very massive, with masses of the order of $100M_\odot$, given the limited cooling of molecular hydrogen \citep[e.g.,][]{bro04}.
This depends on fragmentation in a cosmological minihalo, ionization prior to the onset of gravitational collapse, and the accretion rate from the cloud envelope \citep{mck04,ohk09}, and it seems possible to form lower-mass stars ($\sim10-40M_\odot$) from primordial gas in recent numerical simulations \citep{yos08,sta09}.
The primordial stars with initial masses of $\sim 140-270M_\odot$ explode as pair instability supernovae (PISNe) \citep{barkat,ume02,heg05}.
In terms of chemical abundances, however, no observational signature for the existence of PISNe has been detected.

Chemical enrichment from the first generation of stars has mainly been studied with extremely metal-poor (EMP) stars \citep[e.g.,][]{bee05}.
$10-25$\% of EMP stars with [Fe/H] $\ltsim -2$ \citep{aok10} show carbon enhancement relative to iron ([C/Fe] $\gtsim1$); this is also the case for the three hyper/ultra metal-poor stars known, with [Fe/H] $<-4.5$.
Such stars are collectively known as carbon-enhanced metal-poor (CEMP) stars, and are further subdivided as CEMP-s and CEMP-no stars depending on whether they exhibit or not an enhancement in the abundances of slow neutron capture elements such as Barium.
The abundance patterns of EMP stars are explained with one of, or the combination of, four enrichment sources:
(1) core-collapse supernovae \citep{ume03,iwa05,tom07}, (2) rotating massive stars \citep{mey06}, (3) asymptotic giant blanch (AGB) stars in binary systems \citep{sud04}, or (4) interstellar accretion \citep{yos81,ibe83}.
These enrichment sources have distinct signatures, so that it should be possible to distinguish between them by comparing the observed elemental abundances with theoretical calculations of nucleosynthetic yields.

The high [C/Fe] can be explained with the mass transfer from AGB stars in binary systems, the mass loss from rotating massive stars, and a single core-collapse supernova forming a black hole.
The Ba enhancement can be explained by AGB stars and possibly by rotating massive stars \citep{pig08}.
However, iron-peak elements have to come from supernovae.
The effect of interstellar accretion has not been studied with hydrodynamical simulations, but is observationally estimated to be negligible \citep{fre09}.
The binarity of CEMP stars has been studied by monitoring radial velocity variations.
A signature of binarity is seen statistically for possibly all CEMP-s stars \citep{luc05}, but the binary fraction of CEMP-no stars seems to be much lower \citep{aok10}. In particular, the binarity is not seen for the three stars with [Fe/H] $<-4.5$.
\citet{ume03} were the first to show that the observed abundance pattern from C to Zn 
can be well reproduced with the enrichment from a single core-collapse supernova that leaves behind a relatively massive black hole,
under the assumption of inhomogeneous chemical enrichment \citep{aud95}.
\citet{iwa05} showed that the N abundance can be as large as observed with enhanced mixing between H and He layers during the hydrostatic stellar evolution.

The observations of very metal-poor Damped Lyman $\alpha$ (DLA) systems have opened a new window to study the chemical enrichment of the Universe by the first generations of stars.
DLAs are quasar absorbers defined by their high 
column density of neutral hydrogen, 
$\log N$({H}\,{\sc i})/cm$^{-2} \ge 20.3$.
They appear to sample a range of galaxy types, from the extended 
H\,{\sc i} disks of galaxies, to smaller subgalactic size haloes, 
as well as smaller H\,{\sc i} clouds within larger galaxies \citep{wol05}.
Large scale surveys, such as the Sloan Digital Sky Survey, have increased
ten-fold the number of known DLAs, which now number 
in excess of $\sim 1000$ \citep{not09,pro09}.
Follow-up high resolution spectroscopy of the most metal-poor DLAs is
of particular interest, since the gas they trace 
may have been enriched by very few generations of stars 
\citep{pet08,pen10}.
Moreover, measuring elemental abundances in 
DLAs is straightforward;
the only potential complications are
line saturation and dust depletion and both effects are of 
much reduced importance for metallicities $Z \ltsim 1/100 Z_\odot$.
In addition, the chemical evolution of such systems is rather simple,
whereas in more chemically evolved systems, there are uncertainties 
in the star formation history, gas inflow and outflow, large
contributions from AGB stars and Type Ia supernovae, 
with the result that the signatures of the first stars can be easily washed out.

Thus, if the most metal-poor DLAs 
hold the key to unravelling 
the chemical enrichment from the first stars,
it is of great interest to compare their observed abundance patterns with the nucleosynthesis yields of metal-free stars.
\citet{coo10b} recently reported such a DLA with [Fe/H] $\simeq -3$, which exhibits a strong carbon enhancement relative to all other available elements, including [C/Fe] $\simeq +1.53$.
This reminds us of the CEMP stars in the solar neighborhood.
In this letter, we compare the elemental abundance pattern of the C-rich DLA with the nucleosynthesis yields of both core-collapse and pair-instability supernovae (\S2).
In \S3 we give a more general discussion of
the chemical enrichment of the Universe by the first generation of stars.
 We summarize our main conclusions in \S4.

\section{Abundance Profiling}

In stars with initial masses of $\gtsim 10 M_\odot$, the iron core undergoes gravitational collapse at the end of the star's life. If sufficient amount of the released gravitational energy is transported to the outgoing shock wave, a supernova explosion occurs.
The mechanism of core-collapse explosion and formation of a black hole remnant is still uncertain, although a few groups have presented feasible calculations of exploding supernovae \citep[e.g.,]{mar09}.
Thus, \citet{ume02}, \citet{iwa05}, and \citet{tom07} have calculated nucleosynthesis yields by promptly exploding the progenitor model without calculating further collapse and bounce, and by applying mixing-fallback.
During the supernova explosion, the elements synthesized in different stellar layers mix and a large fraction of this mixed material falls back onto the remnant black hole.
Physically, such a mixing-fallback process can be caused not only by the Rayleigh-Taylor instability \citep{hac90,jog09} but also by a jet-induced explosion \citep{tom09}.
In the present study, we calculate nucleosynthesis yields with two new models of faint core-collapse supernovae applying different mixing-fallback parameters.

\begin{figure}
\center 
\includegraphics[width=8.7cm]{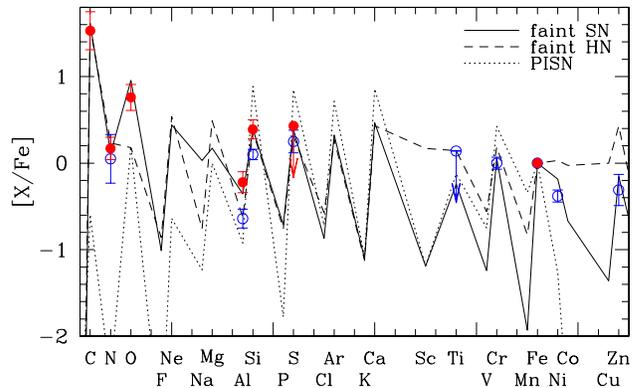}
\caption{\label{fig:dla}
The elemental abundance pattern of the metal-poor C-rich DLA (filled circles) and peculiar DLA (open circles).
The solid and short-dashed lines show the nucleosynthesis yields of faint core-collapse supernovae from $25M_\odot$ stars with mixing-fallback.
The dotted line is for pair-instability supernovae from $170 M_\odot$ stars.
}
\end{figure}

Figure \ref{fig:dla} shows their elemental abundance ratios from C to Zn relative to Fe.
We adopt the progenitor star model from \citet{iwa05} for an initial mass of $M=25M_\odot$ and metallicity $Z=0$.
We implement two cases with different explosion energies: $1 \times 10^{51}$ erg (supernova (SN), solid line) and $20 \times 10^{51}$ erg (hypernova (HN), short-dashed line).
An efficient mixing-fallback is adopted in both models:
the mixing region of ejecta is assumed to be $M_{\rm in}-M_{\rm out}=1.76-6.14M_\odot$ and $2.20-6.30M_\odot$ with the ejection fraction of $f=0.008$ and $0.004$, respectively.
The resultant black hole masses are $6.1M_\odot$ and $6.3M_\odot$, respectively.
The ejected Fe masses are as small as $M(^{56}{\rm Ni})=M({\rm Fe})=0.0018 M_\odot$ and $0.0014 M_\odot$, respectively, and thus correspond to faint supernovae such as SN1997D.
For comparison, $M({\rm Fe})\simeq 0.07 M_\odot$ for normal supernovae and $\gtsim 0.07 M_\odot$ for normal hypernovae in \citet[hereafter K06,][]{kob06}.
The parameter dependencies of elemental abundance ratios can be summarized as follows.
\begin{itemize}
\item
The abundances of odd-Z elements are lower than those of the $\alpha$ elements.
This odd-even effect with the order of $\sim 0.5$ dex is realized for primordial supernovae in general, and is consistent with observations of EMP stars in the Galactic halo \citep[e.g.,][]{cay04}.
The yields of odd-Z elements are enhanced by the surplus of neutrons in $^{22}$Ne, and $^{22}$Ne is transformed from $^{14}$N during He-burning. The absence of seed CNO elements in the progenitor stars results in smaller amounts of odd-Z elements.
This metallicity effect gives excellent agreement with the decreasing trends of [(Na,Al,Cu)/Fe] toward lower metallicity in the solar neighborhood (K06).
The metallicity effect is stronger for Na, Al, Sc, and Cu than for P, Cl, K, V, Mn, and Co.
\item
C and N are synthesized in the outermost region of the ejecta.
Thus, the [C/Fe] ratio is higher for smaller ejection of Fe.
[N/Fe] depends on the initial metallicity and mixing during the stellar evolution.
\item
$\alpha$ elements such as O, Mg, Si, and S are synthesized in the outer regions of the ejecta.
[O/Fe] depends both on the explosion energy and mixing-fallback, and thus it is not possible to put a constraint on the explosion energy only from [O/Fe].
Since Mg is more sensitive to the explosion energy, [Mg/Fe] may give a clue to solve this degeneracy.
Si and S are synthesized in more inner regions than O, and significant amounts of Si and S fallback.
\item
Zn, Co, and V are synthesized in the deep complete Si-burning region together with $\sim 80-90$\% of Fe.
[(Co,Zn)/Fe] $\gtsim 0$ are realized in high energy explosions with mixing-fallback \citep{ume02}; neutrino interactions could enhance [Zn/Fe] \citep{pru05,fro06} but not [Co/Fe] \citep{izu10}.
\item
Cr and Mn are mainly synthesized in the incomplete Si-burning region.
[Cr/Fe] does not vary very much for these models; 
Cr abundance does not depend very much on parameters such as the electron fraction and entropy, and [Cr/Fe] cannot be smaller with smaller $M_{\rm in}$ or larger $M_{\rm out}$.
Larger $M_{\rm in}$ can give larger [Cr/Fe] and smaller [(Co,Zn)/Fe], but such abundance pattern is not consistent with the EMP observations.
Therefore, in galactic chemical evolution models, [Cr/Fe] is almost constant over a wide range of [Fe/H].
This has been a problem when comparing with the observations \citep[e.g.,][]{cay04}, but now is consistent if NLTE effects are taken into account, as shown in K06.
\end{itemize}

We now compare these model calculations with the available abundance measurements 
in the extremely metal-poor DLA reported by \citet[filled circles]{coo10b}.
This DLA was originally identified in the Sloan Digital Sky Survey (SDSS) spectrum
of the QSO J0035$-$0918. Follow-up high resolution spectroscopy indicated
 $z_{\rm abs}=2.3400972$, $\log N({\rm HI})$ $/{\rm cm}^{-2}=20.55 \pm 0.1$,
 [Fe/H] $\simeq -3.04$, and a pronounced carbon enhancement [C/Fe]\,$\simeq +1.53$.
\begin{itemize}
\item The observed [C/Fe] ratio is well reproduced with our faint SN/HN models.
Low-mass AGB stars ($1-4M_\odot$) can also provide such high [C/Fe] \citep[e.g.,][]{kar10}.
However, such low-mass stars are unlikely to contribute at the redshift of the C-rich DLA.
With the star formation history in the solar neighborhood, [C/Fe] reaches the maximum value at $z=1.8$ due to the AGB contribution \citep{kob10}.
\item The [O/Fe] ratio is higher than for non-faint supernovae ([O/Fe] $\sim 0.5-0.6$ for $25M_\odot$ in K06) and is consistent with the faint SN model because of the smaller ejected Fe mass.
\item The [(Si,S)/Fe] ratios are similar to those of non-faint supernovae, and also consistent with the faint SN/HN models.
\item The low Al abundance strongly suggests that the enrichment source is not Pop II supernovae but primordial supernovae.
Even for $Z=0.001$,
[Al/Fe] is larger than $0.35$ for $\ge 20M_\odot$ (K06).
\item The low N abundance is also consistent with the faint SN/HN models, and the high [C/N] ratio cannot be explained with mass loss from rotating massive stars \citep{mey10} or intermediate-mass AGB stars ($\gtsim 4M_\odot$).
\end{itemize}

In order to discuss the detailed explosion mechanism of the supernova, it is necessary to obtain the elemental abundances of iron-peak elements.
For the C-rich DLA, because of the low metallicity, it is impossible to detect heavier elements than S except for Fe.
However, the abundance profiling approach can be used for other absorption
systems with peculiar abundance patterns, because such systems have presumably 
been enriched by only a small number of supernovae.
In Figure \ref{fig:dla} we overplot (open circles) the element abundances recently reported by \citet{coo10} for the $z_{\rm abs}=1.62650$
DLA in front of the gravitationally lensed quasar UM637A.
This DLA, which has $\log N({\rm HI})/{\rm cm}^{-2}=20.7 \pm 0.1$ and [Fe/H] $= -1.56 \pm 0.03$, exhibits a different abundance pattern from other DLAs with similar metallicities and also from the average population of Galactic metal-poor stars.
In particular, Ti, Ni, and Zn are deficient relative to Fe
(see Fig. 11 and 12 in \citealt{coo10}).

We show that this abundance pattern can also be explained with faint supernovae.
The normal [(Si,S)/Fe] ratios are consistent with the supernova scenario.
The low Al abundance strongly suggests that the enrichment source is primordial supernovae.
The observed [Zn/Fe] ratio is more consistent with the faint SN model (solid line) than the faint HN model (dashed line) or non-faint supernovae without mixing-fallback ([Zn/Fe] $\sim -1.7$ in K06). This is also supported by the observed [Ni/Fe]. 
[Cr/Fe] is consistent with both supernova models.
We should note that there is a problem in the Ti nucleosynthesis yields (K06), and we do not include Ti in the abundance profiling.
At this metallicity, however, there is no information on C enhancement in the DLA observations since C line is saturated.
Therefore, it is uncertain how efficient the mixing-fallback is.
In fact, the observed abundances can be explained with medium mixing-fallback of non-faint supernovae.
In this case, the ejected Fe mass is larger, which may be consistent with the observed metallicity, and thus [C/Fe] is lower than for the C-rich DLA.

In Figure \ref{fig:dla}
the dotted line is for the nucleosynthesis yields of the PISN of a $170M_\odot$ star, which are taken from \citet{ume02}.
The primordial stars with initial masses of $\sim 140-270 M_\odot$ enter into the electron-positron pair instability region during the central oxygen-burning stages and contract quasi-dynamically. Then the central temperature increases, central oxygen burning takes place explosively, and the generated nuclear energy is large enough to disrupt the stars completely without leaving compact remnants.
Compared with core-collapse supernovae, the abundance pattern of PISNe can be summarized as follow.
\begin{itemize}
\item The odd-Z effect is much larger than $\sim 1$ dex.
\item ${\rm [(Si,S,Ar,Ca)/Fe]}$ are much larger than [(O,Mg)/Fe] because of more extensive explosive oxygen burning.
\item ${\rm [Cr/Fe]}$ is much larger because of the larger incomplete Si-burning region.
\item ${\rm [(Co,Zn)/Fe]}$ are much smaller because of the much larger ratio between the complete and incomplete Si-burning regions.
\end{itemize}
All of these characteristics disagree with the observed elemental abundances of these metal-poor DLAs.
Even at high-redshift, there is no signature of the existence of PISNe.
The [Si/C] for PISNe is as large as $+1.5$, which is also inconsistent with the observational estimate in the intergalactic medium (IGM) ([Si/C] $\sim 0.77$, \citealt{agu04}).
The IGM abundance looks more consistent with normal (non-faint) core-collapse supernovae with [C/Fe] $\sim 0$ and [Si/Fe] $\sim 0.7$ (K06).

\section{Discussion}

We have shown that the enrichment source of the extremely metal-poor DLA
is very likely a primordial supernova that is faint as a result of mixing-fallback to form a $3-6M_\odot$ black hole.
It is interesting that the observed DLA abundance is very similar to those of EMP stars in the solar neighborhood including the ultra metal-poor star HE0557-4840 ([Fe/H] $=-4.75$, [C/Fe] $=+1.6$, \citealt{nor07}) and BD+44$^\circ$493 \citep{ito09}.
Without rotation, the chemical enrichment from very massive stars ($\gtsim 50-100 M_\odot$) is very small because the central part is supposed to fall onto the black hole.
In the early stages of galaxy formation, chemical enrichment is likely to be driven by core-collapse supernovae from $\sim 20-50 M_\odot$ stars, although the mass range depends on the rotation.

How did the first supernova enrich the first galaxy?
In our supernova scenario, we assume that the first supernova occurs in a primordial gas cloud with the total mass of $10^{6-7} M_\odot$.
In hydrodynamical simulations \citep[e.g.,][]{mac99}, the interstellar medium (ISM) is ionized by the supernova explosion, and the HI mass rapidly decreases. Afterwards, the HI mass slightly increases due to the recombination, and then decreases due to galactic mass-outflow.
Some of the ISM could have the HI column density of $\log N({\rm HI})\sim 20.5$ and could be observed as DLAs.
The observable HI regions could have $M({\rm HI})$ $\sim 3000M_\odot$, although the HI mass highly depends on the total mass of the galaxy, the radial density profile, and the inhomogeneity of the ISM.
With our faint supernova models, the ejected C mass is $\sim 0.2M_\odot$,
which is roughly consistent with the observed C-rich DLA.
In \citet{coo10b}, the masses of carbon and neutral gas of the C-rich DLA are estimated as
$M({\rm CII}) \sim 2 \left(n({\rm H})/1\,{\rm cm}^{-3}\right)^{-2} M_\odot$
and
$M({\rm HI}) \sim 2.5 \times 10^4 \left(n({\rm H})/1\,{\rm cm}^{-3}\right)^{-2} M_\odot$,
which are consistent with our scenario if $n({\rm H}) \sim 3\,{\rm cm}^{-3}$.

Are there any signatures of the first stars in other metal-poor systems?
For globular cluster systems (GCSs), the present stellar mass and half-light radius are $10^{4-6}M_\odot$ and $1-35$ pc \citep[e.g.,][]{gil07}, respectively, which imply high densities ($n \sim 1000$).
Since the contribution of dark matter is small, the total mass would be $10^{5-9} M_\odot$ with the range of star formation efficiencies ($0.001-0.1$).
The progenitors of GCSs could be more massive if the GCS have lost a large fraction of stars by the relaxation as shown in N-body simulations \citep{lam10}.
Star formation takes place quickly, and $\gtsim 10-1000$ supernovae are expected to occur from the stellar mass with the Salpeter initial mass function, which might blow away the ISM and quench star formation.
This is consistent with the narrow metallicity distribution functions and the lack of the scatter in elemental abundance ratios \citep{car09}.
In the elemental abundance patterns, no carbon enhancement that seems to be originated from supernovae is seen.

For dwarf spheroidal galaxies (dSphs), the present stellar mass and half-light radius are $10^{3-7}M_\odot$ and $20-1000$ pc \citep[e.g.,][]{gil07}, respectively, 
implying much lower densities than in GCSs.
The dark matter mass is about $10^7M_\odot$ independent of the stellar mass \citep{geh09}. 
From the stellar mass, $1-10^4$ supernovae are expected.
As seen in the observed color-magnitude diagrams \citep{tol09}, 
star formation takes place slowly with a very low rate, and thus the ISM is likely to be inhomogeneous.
The elemental abundance pattern can be used for the abundance profiling.
In fact, a few EMP stars at [Fe/H] $\ltsim -3.5$ in dSphs do show carbon enhancement \citep{nor10}, which suggests a contribution from faint supernovae.
Also in the outer halo of the Milky Way Galaxy, three stars at [Fe/H] $\ltsim -2.5$ do show [C/Fe] $\sim +2$ (Beers 2010, private communication);
 such stars may be disrupted from dSphs.

\section{Conclusions}

In the early stages of chemical enrichment, the interstellar medium is supposed to be highly inhomogeneous, so that the properties of the first objects can be directly extracted from the comparison between the observed elemental abundances and nucleosynthesis yields.
We have shown that the observed abundance pattern of the very metal-poor C-rich DLA 
is in excellent agreement with the nucleosynthesis yields of a 
primordial star that explodes as a faint core-collapse supernova owing to the efficient mixing and fallback.
The nucleosynthesis yields of PISNe are not consistent with the observation.
The contribution from rotating massive stars seems to be small because of the lack of N enhancement.
The contribution from AGB stars should be very small because of the N abundance and of the enrichment timescale.
Since the DLA abundances reflect the chemical enrichment in gas-phase, the binary or accretion scenarios of the EMP stars do not work.
Thus,
we conclude that enrichment by primordial supernovae is the best solution to explain the abundance pattern of the C-rich DLA.
The abundance pattern of the C-rich DLA is similar to those of EMP stars such as the ultra metal-poor star HE0557-4840.
Some of EMP stars in dSphs and the Galactic outer halo also show similar carbon enhancement at [Fe/H] $\ltsim -3$.
Chemical enrichment by the first stars in the first galaxies is likely to be driven by core-collapse supernovae.

\acknowledgments
We would like to thank M. Pettini and R. Cooke for providing their results prior to publication.
We also thank J. Norris, R. Sutherland, G. Da Costa, D. Yong, and D. Mackey for fruitful discussion.
This work has been supported in part by WPI Initiative, MEXT, Japan, and by the Grant-in-Aid for Scientific Research of the JSPS (18104003, 20540226) and MEXT (19047004, 22012003).

\end{document}